\documentclass[
    ,final            
  ]
  {aipproc}

\layoutstyle{6x9}
\newcommand{\ba}{\begin{eqnarray}}
\newcommand{\ea}{\end{eqnarray}}

\begin{document}

\title{Generalized Partial Dynamical Symmetries in~Nuclear Spectroscopy}

\author{A. Leviatan}{
  address={Racah Institute of Physics, The Hebrew University,
Jerusalem 91904, Israel}
}



\begin{abstract}
Explicit forms of IBM Hamiltonians with a generalized partial dynamical 
O(6) symmetry are presented and compared with empirical data in $^{162}$Dy.
\end{abstract}

\maketitle



A dynamical symmetry corresponds to a situation in which the Hamiltonian
is written in terms of the Casimir operators of a chain of nested algebras
\ba
G_1 \supset G_2 \supset \ldots \supset G_n ~,
\label{ds}
\ea
and has the following properties. (i) Solvability. 
(ii) Quantum numbers related to irreducible representations (irreps) of the 
algebras in the chain. (iii) Symmetry-dictated structure of wave functions 
independent of the Hamiltonian's parameters.
The merits of a dynamical symmetry are self-evident, 
however, in most applications to realistic systems, 
one is compelled to break it. Partial dynamical symmetry (PDS) 
corresponds to a particular symmetry breaking for which some (but not all) 
of the above virtues of a dynamical symmetry are retained.
Two types of partial symmetries were encountered so far. 
The first type correspond to a situation for which
{\bf part} of the states preserve {\bf all} the dynamical symmetry.
This is the case for the SU(3) PDS found in the IBM-1 
\cite{lev96,levsin99} and the 
Symplectic Shell Model \cite{esclev00,esclev02}, 
and for the $F$-spin PDS in the IBM-2 \cite{levgin00}. 
The corresponding PDS Hamiltonians have a subset 
of solvable states with good symmetry while other eigenstates are mixed.
A second type of partial
symmetries correspond to a situation for which {\bf all} the
states preserve {\bf part} of the dynamical symmetry.
This occurs, for example, when the Hamiltonian
preserves only some of the symmetries $G_i$ in the chain (\ref{ds}) and
only their irreps are unmixed \cite{talmi97,isa99}.
In this case there are no analytic solutions, yet selected quantum 
numbers (of the conserved symmetries) are retained. 
In the present contribution we show that it is possible to combine both 
types of partial symmetries, namely, to construct a Hamiltonian for 
which {\bf part} of the states have {\bf part} of the dynamical symmetry. 
We refer to such a structure as a generalized partial dynamical symmetry 
\cite{levisa02}.

Partial symmetry of the second kind was recently
considered in \cite{isa99} in relation to the chain
\ba
U(6) \supset O(6) \supset O(5) \supset O(3) ~.
\label{dsO6}
\ea
The Hamiltonian employed has two- and three-body interactions of the form
\ba
H_1 \;=\; \kappa_{0}P^{\dagger}_{0}P_{0}
+ \kappa_2 \Bigl (\Pi^{(2)}\times \Pi^{(2)}\Bigr )^{(2)}\cdot\Pi^{(2)} ~.
\label{h1}
\ea
The $\kappa_0$ term is the $O(6)$ pairing term defined in terms of
monopole ($s$) and quadrupole ($d$) bosons,
$P^{\dagger}_{0}=  d^{\dagger}\cdot d^{\dagger} - (s^{\dagger})^2$.
It is diagonal in the dynamical symmetry basis
$\vert [N],\sigma,\tau,L\rangle$ of Eq. (\ref{dsO6}) with eigenvalues
$\kappa_0(N - \sigma)(N +\sigma +4)$.
The $\kappa_2$ term is composed only of the $O(6)$ generator: 
$\Pi^{(2)}=d^{\dagger}s+s^{\dagger}\tilde{d}$, 
which is not a generator of $O(5)$. 
Consequently, $H_1$ cannot connect different $O(6)$ irreps but
can induce $O(5)$ mixing. The eigenstates have good $\sigma$ but not good 
$\tau$ quantum numbers.

To consider a generalized $O(6)$ PDS, we introduce the
following IBM-1 Hamiltonian,
\ba
H_2 \;=\; h_{0}P^{\dagger}_{0}P_{0} + h_{2}P^{\dagger}_{2}
\cdot\tilde P_{2} ~.
\label{h2}
\ea
The $h_0$ term is identical to the $\kappa_0$ term of Eq. (\ref{h1}), and
the $h_2$ term is defined in terms of the boson pair
$P^{\dagger}_{2,\mu} = \sqrt{2}\,s^{\dagger}d^{\dagger}_{\mu}
+ \sqrt{7}(d^{\dagger}d^{\dagger})^{(2)}_{\mu}$ with
$\tilde P_{2,\mu} = (-)^{\mu}P_{2,-\mu}$. The latter term can induce 
both $O(6)$ and $O(5)$ mixing.
Although $H_2$ is not an $O(6)$ scalar, it has an exactly
solvable ground band with good $O(6)$ symmetry. This
arises from the fact that the $O(6)$ intrinsic state for the
ground band
\ba
\vert c;\,N \rangle = (N!)^{-1/2}(b^{\dagger}_{c})^{N}\vert 0 \rangle
\;, \;\;\;
b^{\dagger}_c = (d^{\dagger}_{0} +s^{\dagger} )/\sqrt{2} ~,
\label{cond}
\ea
has $\sigma=N$ and is an exact zero energy eigenstate of $H_2$.
Since $H_2$ is rotational invariant, states of good angular momentum $L$
projected from $\vert c;\,N\rangle$ are also zero-energy eigenstates of 
$H_2$ with good $O(6)$ symmetry, and form the ground band of 
$H_2$. 
It follows that $H_2$ 
has a subset of solvable states with good $O(6)$ symmetry 
($\sigma=N$), which is not preserved by other states. 
All eigenstates of $H_2$ break the $O(5)$ symmetry but preserve the 
$O(3)$ symmetry. These are precisely the required features of a generalized 
partial dynamical symmetry as defined above 
for the chain of Eq.~(\ref{dsO6}).  

\begin{figure}
   \includegraphics[angle=270,totalheight=10cm]{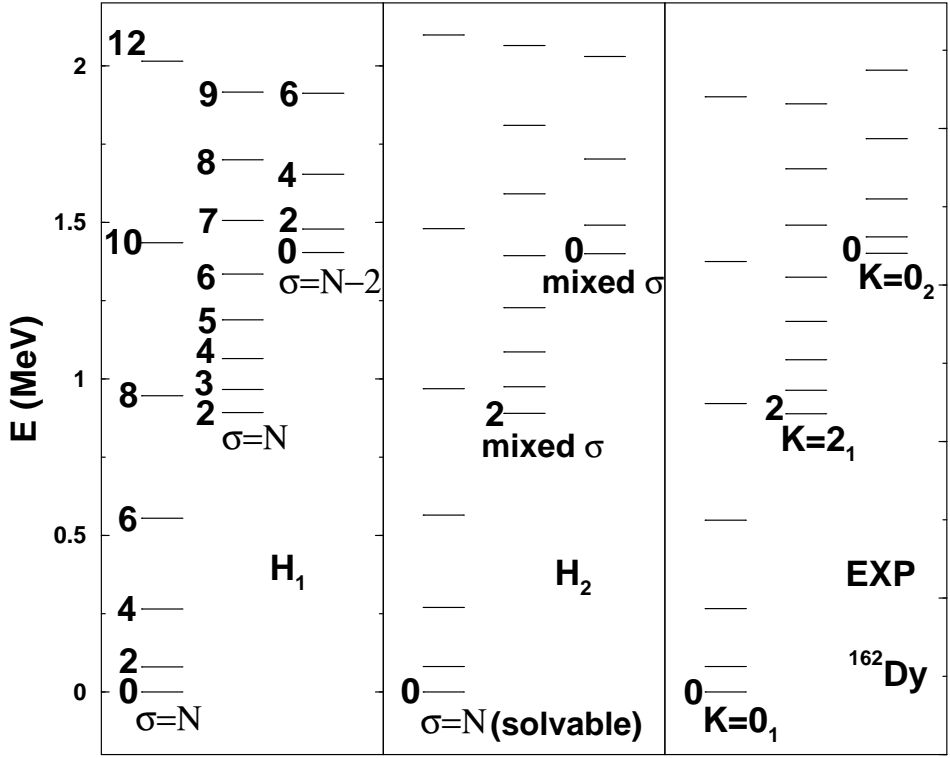}
  \caption{Experimental spectra (EXP) of $^{162}$Dy [9,10] 
compared with calculated spectra  
of $H_1 +\lambda_{1} L\cdot~L$, Eq.~(\ref{h1}), and 
$H_2+\lambda_2 L\cdot L$, Eq.~(\ref{h2}), with parameters
$\kappa_0=8$, $\kappa_2=1.364$, $\lambda_1=8$, and 
$h_0=28.5$, $h_2=6.3$, $\lambda_2=13.45$ keV and $N=15$.}
\end{figure}
In Fig.~1 we show the experimental spectrum 
of $^{162}$Dy and compare with the calculated spectra of $H_1$ and $H_2$. 
The spectra display rotational bands of an axially-deformed nucleus, 
in particular, a ground band $(K=0_1)$ and excited $K=2_1$ and 
$K=0_2$ bands. An $L\cdot L$ term was added to both Hamiltonians, which 
contributes to the rotational splitting but has no effect on wave 
functions. The parameters were chosen to reproduce the excitation 
energies of the $2^{+}_{K=0_1},\, 2^{+}_{K=2_1}$ and $0^{+}_{K=0_2}$ 
levels. 
The $O(6)$ decomposition of selected bands is shown in Fig.~2. 
For $H_2$, the solvable $K=0_1$ ground band has $\sigma=N$ and 
exhibits an exact $L(L+1)$ splitting. 
The $K=2_1$ band is almost pure with only $0.15\%$ admixture of 
$\sigma=N-2$ into the dominant $\sigma=N$ component. The $K=0_2$ band 
has components with $\sigma=N\,(85.50\%)$, $\sigma=N-2\,(14.45\%)$, 
and $\sigma=N-4\,(0.05\%)$. Higher bands exhibit stronger mixing, 
{\it e.g.}, the $K=2_3$ band shown in Fig.~2, has components 
with $\sigma=N\,(50.36\%)$, $\sigma=N-2\,(49.25\%)$, 
$\sigma=N-4\,(0.38\%)$, and $\sigma=N-6\,(0.01\%)$.
\begin{figure}
   \includegraphics[angle=270,totalheight=10cm]{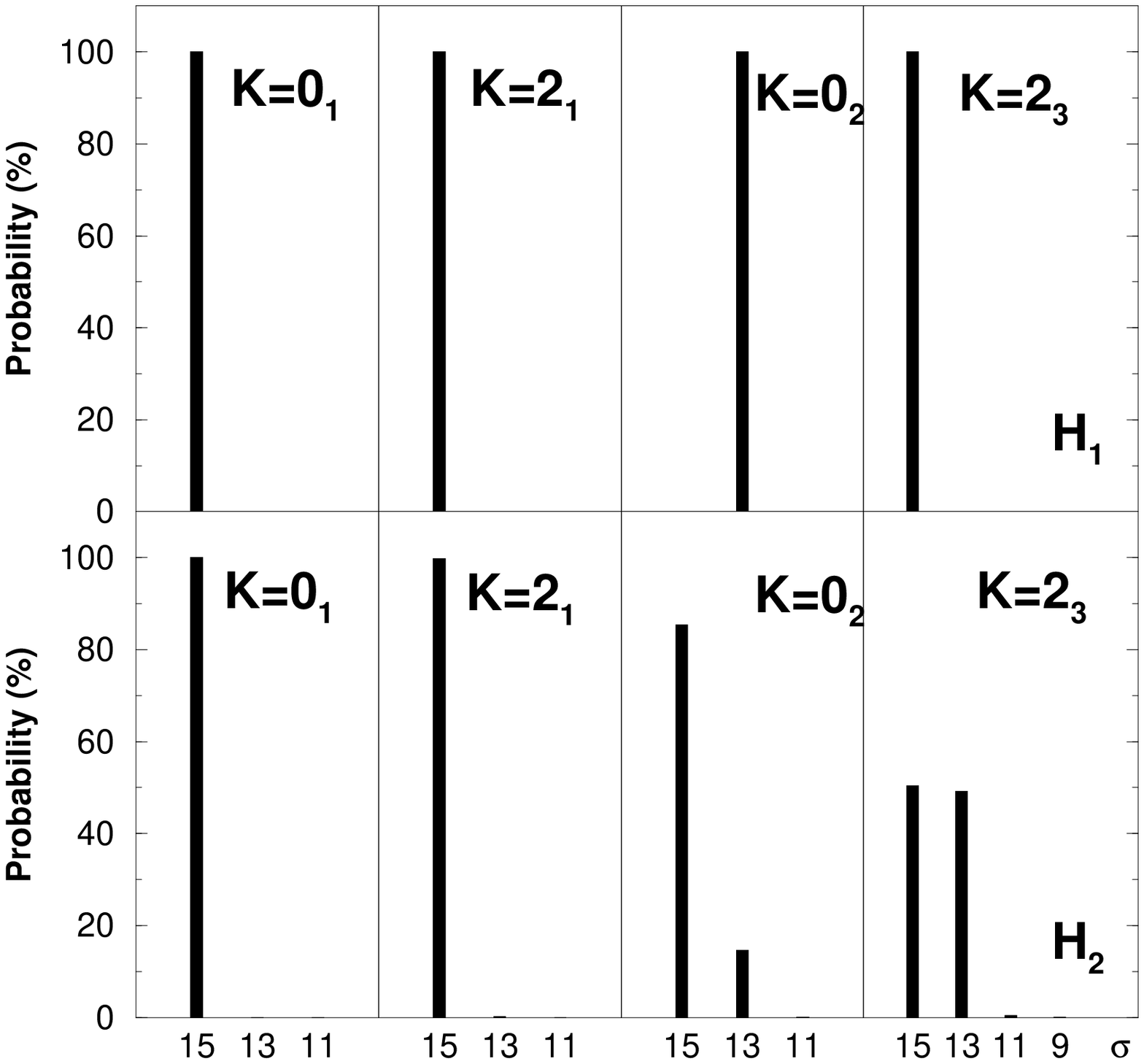}
  \caption{$O(6)$ decomposition of wave functions of the 
$K=0_1,\,2_1,\,0_2,\,2_3$ bands for $H_1$ (upper portion) and 
$H_2$ (lower portion).}
\end{figure}
The $O(6)$ mixing in excited bands of $H_2$ 
depends critically on the ratio $h_2/h_0$ in Eq.~(\ref{h2}) or equivalently 
on the ratio of the $K=0_2$ and $K=2_1$ bandhead energies.
In contrast, all bands of $H_1$ are pure with respect to $O(6)$.
Specifically, the $K=0_1,2_1,2_3$ bands shown in Fig.~2 have 
$\sigma=N$ and the $K=0_2$ band has $\sigma=N-2$. 
In this case the diagonal $\kappa_0$ term in Eq.~(\ref{h1}) simply 
shifts each band as a whole in accord with its $\sigma$ assignment. 
All eigenstates of both $H_1$ and $H_2$ are mixed 
with respect to $O(5)$. 

To gain more insight into the underlying band structure of $H_2$ we 
perform a band-mixing calculation by taking its matrix elements  
between large-N intrinsic states. 
The latter are obtained in the usual 
way by replacing a condensate boson in 
$\vert c;\,N \rangle$ (\ref{cond}) with orthogonal bosons 
$b^{\dagger}_{\beta}=(d^{\dagger}_0 - s^{\dagger})/\sqrt{2}$ and 
$d^{\dagger}_{\pm 2}$ representing $\beta$ and $\gamma$ excitations 
respectively. By construction, the intrinsic state 
for the ground band of $H_2$, $\vert K=0_1 \rangle = \vert c;\, N\rangle$, 
is decoupled. For the lowest excited bands we find
\ba
\vert K=0_2\, \rangle &=& 
A_{\beta}\,\vert\beta\rangle +
A_{\gamma^2}\,\vert\gamma^2_{K=0}\,\rangle +
A_{\beta^2}\,\vert\beta^2\,\rangle ~,
\nonumber\\
\vert K=2_1\, \rangle &=& 
A_{\gamma}\,\vert\gamma\,\rangle +
A_{\beta\gamma}\,\vert\beta\gamma\,\rangle ~.
\label{intstat}
\ea
Using the parameters of $H_2$ relevant to $^{162}$Dy (see Fig.~1) 
we obtain that the $K=0_2$ band is composed of   
$36.29\%$ $\beta$, $63.68\%$ $\gamma^2_{K=0}$ and 
$0.03\%$ $\beta^2$ modes, {\it i.e.}, it is dominantly a double-gamma 
phonon excitation with significant single-$\beta$ phonon admixture. 
The $K=2_1$ band is composed of 
$99.85\%$ $\gamma$ and $0.15\%$ $\beta\gamma$ modes, {\it i.e.}
it is an almost pure single-gamma phonon band. An $O(6)$ decomposition 
of the intrinsic states in Eq.~(\ref{intstat}) shows that 
the $K=0_2$ intrinsic state has components with $\sigma=N\,(86.72\%)$, 
$\sigma=N-2\,(13.26\%)$ and $\sigma=N-4\,(0.02\%)$. 
The $K=2_1$ intrinsic state has $\sigma=N\,(99.88\%)$ and 
$\sigma=N-2\,(0.12\%)$. These estimates are in good agreement 
with the exact results mentioned above in relation to Fig.~2.

\begin{table}
\begin{tabular}{llll|llll}
\hline
\tablehead{1}{l}{b}{Transition} & 
\tablehead{1}{l}{b}{$\mathbf{H_{1}}$} & 
\tablehead{1}{l}{b}{$\mathbf{H_{2}}$} & 
\tablehead{1}{l}{b}{Expt.} &
\tablehead{1}{l}{b}{Transition} & 
\tablehead{1}{l}{b}{$\mathbf{H_{1}}$} & 
\tablehead{1}{l}{b}{$\mathbf{H_{2}}$} & 
\tablehead{1}{l}{b}{Expt.} \\
\hline
$2^{+}_{K=0_1}\rightarrow 0^{+}_{K=0_1}$  & 1.06   & 1.05   & 1.07(2) &
$2^{+}_{K=2_1}\rightarrow 0^{+}_{K=0_1}$  & 0.024  & 0.024  & 0.024(1)  \\
$4^{+}_{K=0_1}\rightarrow 2^{+}_{K=0_1}$  & 1.50   & 1.49   & 1.51(6) &
$2^{+}_{K=2_1}\rightarrow 2^{+}_{K=0_1}$  & 0.038  & 0.0395 & 0.042(2)  \\
$6^{+}_{K=0_1}\rightarrow 4^{+}_{K=0_1}$  & 1.62   & 1.61   & 1.57(9) &
$2^{+}_{K=2_1}\rightarrow 4^{+}_{K=0_1}$  & 0.0025 & 0.0026 & 0.0030(2) \\
$8^{+}_{K=0_1}\rightarrow 6^{+}_{K=0_1}$  & 1.65   & 1.65 & 1.82(9)   &
$3^{+}_{K=2_1}\rightarrow 2^{+}_{K=0_1}$  & 0.0428 & 0.0425 &           \\
$10^{+}_{K=0_1}\rightarrow 8^{+}_{K=0_1}$ & 1.63   & 1.64 & 1.83(12)  &
$3^{+}_{K=2_1}\rightarrow 4^{+}_{K=0_1}$  & 0.022  & 0.023  &           \\
$12^{+}_{K=0_1}\rightarrow 10^{+}_{K=0_1}$& 1.58   & 1.60 & 1.68(21)  &
$4^{+}_{K=2_1}\rightarrow 2^{+}_{K=0_1}$ & 0.0123 & 0.0114 & 0.0091(5)  \\
 & & & &
$4^{+}_{K=2_1}\rightarrow 4^{+}_{K=0_1}$ & 0.046  & 0.047  & 0.044(3)   \\
$0^{+}_{K=0_2}\rightarrow 2^{+}_{K=0_1}$ & 0.0014  & 0.0022 &         & 
$4^{+}_{K=2_1}\rightarrow 6^{+}_{K=0_1}$ & 0.0061 & 0.0061 & 0.0063(4)  \\
$0^{+}_{K=0_2}\rightarrow 2^{+}_{K=2_1}$ & 0.0012  & 0.1707 &         &
$5^{+}_{K=2_1}\rightarrow 4^{+}_{K=0_1}$ & 0.0345 & 0.033  & 0.033(2)   \\
$2^{+}_{K=0_2}\rightarrow 0^{+}_{K=0_1}$ & 0.0002  & 0.0004 &         &
$5^{+}_{K=2_1}\rightarrow 6^{+}_{K=0_1}$ & 0.029  & 0.031  & 0.040(2)   \\
$2^{+}_{K=0_2}\rightarrow 2^{+}_{K=0_1}$ & 0.0003  & 0.0005 &         &
$6^{+}_{K=2_1}\rightarrow 4^{+}_{K=0_1}$ & 0.0085 & 0.0071 & 0.0063(4)  \\
$2^{+}_{K=0_2}\rightarrow 2^{+}_{K=2_1}$ & 0.0003  & 0.0365 &         &
$6^{+}_{K=2_1}\rightarrow 6^{+}_{K=0_1}$ & 0.046  & 0.047  & 0.050(4)   \\
\hline
\end{tabular}
\caption{Calculated and observed [10,11] $B(E2)$ values 
$(e^2b^2)$ for $^{162}$Dy. The $E2$ parameters in Eq.~(\ref{e2}) 
are $e=0.138$ $(0.126)$ $eb$ and $\chi=-0.22$ $(-0.55)$ 
for $H_1$ $(H_2)$.} 
\end{table}

In Table 1 we compare the presently known experimental $B(E2)$ 
values for transitions in $^{162}$Dy with the values predicted by 
$H_1$ and $H_2$ using the $E2$ operator
\ba
T^{(2)} &=& 
e\left[\, \Pi^{(2)} + \chi\, (d^{\dagger}\tilde{d}\,)^{(2)}\,\right ] ~.
\label{e2}
\ea
The parameters $e$ and $\chi$ in Eq.~(\ref{e2}) were fixed for each 
Hamiltonian by the empirical $2^{+}_{K=0_1}\rightarrow 0^{+}_{K=0_1}$ and 
$2^{+}_{K=2_1}\rightarrow 0^{+}_{K=0_1}$ $E2$ rates. 
The $B(E2)$ values predicted by $H_1$ and $H_2$ for 
$K=0_1\rightarrow K=0_1$ and $K=2_1\rightarrow K=0_1$ transitions  
are very similar and agree well with the measured values. 
On the other hand, their predictions for interband 
transitions from the $K=0_2$ band are very different. 
For $H_1$, the $K=0_2\rightarrow K=0_1$ and $K=0_2\rightarrow K=2_1$ 
transitions are comparable and weaker than $K=2_1\rightarrow K=0_1$. 
This can be understood if we recall the $O(6)$ assignments for the bands 
of $H_1$: $K=0_1,\,2_1\,(\sigma=N)$, $K=0_2\,(\sigma=N-2)$, and the $E2$ 
selection rules of $\Pi^{(2)}\,(\Delta\sigma=0)$ and 
$(d^{\dagger}\tilde{d}\,)^{(2)}\,(\Delta\sigma=0\pm 2)$, which imply 
that in this case only the $(d^{\dagger}\tilde{d}\,)^{(2)}$ term  
contributes to interband transitions from 
the $K=0_1$ band. In contrast, for $H_2$, 
$K=0_2\rightarrow K=2_1$ and $K=2_1\rightarrow K=0_1$ transitions are 
comparable and stronger than $K=0_2\rightarrow K=0_1$. 
This behaviour is due to the underlying band structure discussed above,  
and the fact that 
$\langle K=0_2\, \vert \Pi^{(2)}_{0} \vert\, K=0_1\rangle =0$, 
while both terms in Eq.~(\ref{e2}) contribute to $\Delta K=2$ interband 
$E2$ intrinsic matrix elements. Recently the $B(E2)$ ratios 
$R_1={B(E2;\,0^{+}_{K=0_2}\rightarrow 2^{+}_{K=2_1})\over 
B(E2;\,0^{+}_{K=0_2}\rightarrow 2^{+}_{K=0_1})} = 10(5)$ 
and $R_2={B(E2;\,2^{+}_{K=0_2}\rightarrow 4^{+}_{K=0_1})\over 
B(E2;\,2^{+}_{K=0_2}\rightarrow 0^{+}_{K=0_1})} = 65(28)$ 
have been measured \cite{zamfir99}. 
The corresponding predictions are $R_1= 0.86$, $R_2=4.00$ for $H_1$ 
and $R_1=77.59$, $R_2=3.25$ for $H_2$. 
As noted in \cite{zamfir99}, the empirical value of $R_2$ deviates 
`beyond reasonable expectations' from the Alaga rule value $R_2=2.6$.
A measurement of absolute 
B(E2) values for these transitions is highly desirable to clarify the 
origin of these discrepancies.\\ 


It is a pleasure to dedicate this article to Rick Casten on the occasion 
of his 60th birthday, and thank him for many years of illuminating 
discussions. This work was done in collaboration with P. Van Isacker (GANIL) 
and was supported in part by the Israel Science Foundation.  

\end{document}